\documentstyle[preprint,eqsecnum,aps,amsfonts,amssymb,floats]{revtex}

\input FEYNMAN

\begin{document}

\draft

\preprint{CALT-68-2098}

\title{Perturbative Corrections to a Sum Rule for the \\Heavy Quark Kinetic Energy}

\author{Hooman Davoudiasl and Adam K. Leibovich
\footnote{The work of H.D. and A.K.L. was supported in part by the
U.S. Department of Energy under Grant No. DE-FG03-92-ER40701.}
}

\address{
California Institute of Technology\\ Pasadena, CA 91125 }

\date{\today}

\maketitle

\begin{abstract}
We calculate the perturbative corrections to order $\alpha_s^2\beta_0$
to the sum rule derived from the second moment of the time-ordered
product of $b \to c$ currents near zero recoil.  This sum rule yields
a bound on $\lambda_1$, the expectation value of the $b$ quark kinetic
energy operator inside the $B$ meson.  The perturbative corrections
significantly weaken the bound relative to the tree level result,
yielding $\lambda_1 < -0.15 {\mathrm\ GeV}^2$.
\end{abstract}


\widetext

\section{Introduction}

Heavy Quark Effective Theory (HQET) is a powerful tool for studying
the decays of hadrons containing one heavy quark $Q$.  For $B \to D^{(*)}
e \bar{\nu}_e$ decays all form factors, at leading order
in the $1/m_Q$ expansion, are related to the Isgur--Wise function.
This function is normalized to unity at zero recoil \cite{IsgurWise,Luke}.

Combining HQET and the operator product expansion (OPE) has also led
to improvements in the understanding of inclusive $B$ decays
\cite{CGG}.  It is possible to show, at leading order in $1/m_Q$, that
the inclusive semileptonic $B$ decay rate is equal to the free $b$
quark decay rate.  Corrections to this result enter at order $1/m_Q^2$
\cite{BUV,BSUV,ManoharWise,BKSV,Neubert} and are paramaterized by two
nonperturbative matrix elements,
\begin{equation}\label{lambda1}
\lambda_1 \equiv \frac{1}{2 m_M}\langle M(v)|\bar{h}_v^{(b)} 
(iD)^2 h_v^{(b)}|M(v) \rangle
\end{equation}
and
\begin{equation}\label{lambda2}
\lambda_2 \equiv \frac{1}{2 d_M m_M}\langle M(v)|\bar{h}_v^{(b)} 
\frac{g_s}{2}\sigma_{\mu\nu}G^{\mu\nu} h_v^{(b)}|M(v) \rangle,
\end{equation}
where $h_v^{(b)}$ is the $b$ quark field in HQET, $M$ is either a $B$
or a $B^*$ meson, $d_B = 3$ and $d_{B^*} = -1$.  These parameters, along with
another parameter $\overline{\Lambda}$, also enter into the relation
between the quark pole mass and the hadron mass
\begin{equation}\label{massrelation}
m_M \equiv m_b + \overline{\Lambda} - \frac{\lambda_1 + d_M \lambda_2}{2m_b}
+ {\mathcal O}(1/m_b^2).
\end{equation}
While it is possible to obtain a value for $\lambda_2$ from the
measured $B - B^*$ mass splitting, extractions of $\overline{\Lambda}$
and $\lambda_1$, although well studied, have large uncertainties
\cite{BallBraun,GKLW}.

By taking appropriate moments of the time-ordered product of $b \to c$
currents, it is possible to obtain sum rules that relate the exclusive
decay form factors to the HQET nonperturbative parameters
\cite{BSUV,BGSUV}.  Taking the zeroth moment of the time-ordered
product yields the Bjorken sum rule \cite{Bjorken,IsgurWise2}. This
sum rule bounds the slope parameter $\rho^2 \equiv - d\xi/dw|_{w=1}$
from below, where $\xi(w)$ is the Isgur--Wise function, $w = v\cdot
v'$, $v$ is the four-velocity of the $B$ and $v'$ is the four-velocity
of the $D^{(*)}$.  The first moment can be used to derive the Voloshin
sum rule \cite{Voloshin}, which bounds $\rho^2$ from above.  The
zeroth moment and the second moment can be combined to obtain a bound
relating $\lambda_1$ to $\rho^2$ \cite{BSUV,BGSUV}.

In this paper, we calculate the perturbative QCD corrections to this
third sum rule to ${\mathcal O}(\alpha_s^2 \beta_0)$ and leading order
in $1/m_Q$.  We also consider redefining $\lambda_1$ and
$\overline{\Lambda}$ in order to absorb these corrections, and compare
these redefinitions to those suggested by other previously studied
sum rules \cite{KLWG,BLRW}.  It is of interest to see whether the
perturbative redefinition can be achieved in a universal way,
regardless of the sum rule in question.

Our paper is organized as follows.  In the next section, we introduce
the formalism used in our work.  In section III, we present the
perturbative corrections to the aforementioned third sum rule and
derive a bound on $\lambda_1$.  Section IV contains a discussion of
this bound and the utility of the redefinition of $\lambda_1$ and
$\overline{\Lambda}$.

\section{Formalism}
We  introduce the time-ordered product
\begin{equation}\label{Tmn}
T_{\mu\nu} = \frac{i}{2 m_B}\int d^4x e^{-iq\cdot x}
\langle B|T\{J_\mu^\dag(x),J_\nu(0)\}|B \rangle,
\end{equation}
where $J_\mu$ is a $b \to c$ axial or vector current, $|B\rangle$
represents the $B$ meson state at rest, and $q$ is the four-momentum
transfer. Here the three-momentum transfer $\vec{q}$ is fixed and $q^0
= M_B - E_M-\varepsilon$, where $E_M = \sqrt{m_M^2 + |\vec{q}|^2}$ is
the minimal possible energy of the hadronic final state associated
with $J_\mu$.  The time-ordered product $T_{\mu\nu}$ has two cuts in
the complex $\varepsilon$-plane.  One cut lies along the positive real
axis $0 < \varepsilon < \infty$.  The second cut, corresponding to
physical states with two $b$-quarks and a $\bar{c}$-quark, lies along
$-\infty < \varepsilon < - 2 E_M$.  This second cut does not affect
our results.

By contracting $T_{\mu\nu}$ with an appropriate four-vector $a$, it is
possible to isolate specific hadronic form factors.  Let
\begin{equation}
T(\varepsilon) \equiv a^{*\mu}T_{\mu\nu}(\varepsilon)a^\nu.
\end{equation}
Inserting a complete set of states $X$ between $J_\mu(x)$
and $J_\nu(0)$ in Eq.~(\ref{Tmn}) yields
\begin{equation}\label{Te}
T(\varepsilon) = \frac{1}{2m_B} \sum_X(2\pi)^3\delta^{(3)}(\vec{q} +\vec{p}_X)
\frac{|\langle X|J\cdot a|B\rangle|^2}{E_X - E_M - \varepsilon} + \cdots,
\end{equation}
where the ellipsis represents the contribution from the other cut.  By
integrating over $\varepsilon$, the following zeroth moment sum rule
is obtained
\begin{equation}\label{zeromoment}
\frac{1}{2 \pi i}\int_C {\mathrm d}\varepsilon 
\theta(\varepsilon - \Delta) T(\varepsilon) =
\sum_X \theta(E_X - E_M - \Delta) (2 \pi)^3 \delta^3(\vec{q} + \vec{p}_X)
\frac{\left | \langle X|J\cdot a | B\rangle \right |^2}{2 m_B}.
\end{equation}
In Eq.~(\ref{zeromoment}), we have included a $\theta$-function which
corresponds to summing over all hadronic resonances up to an
excitation energy $\Delta$.  Relating the integral with the hard
cutoff to the exclusive states above requires local duality at the
scale $\Delta$.

The second moment of $T(\varepsilon)$ gives
\begin{eqnarray}\label{secondmoment}
\frac{1}{2 \pi i}\int_C {\mathrm d}\varepsilon\,\varepsilon^2 
\theta(\varepsilon - \Delta) T(\varepsilon) &=&
\sum_{X\neq M} \theta(E_X - E_M - \Delta) (E_X - E_M)^2\nonumber\\
&&\qquad \times  (2 \pi)^3 \delta^3(\vec{q} + \vec{p}_X)
\frac{\left | \langle X|J\cdot a | B\rangle \right |^2}{2 m_B}.
\end{eqnarray}
A combination of the zeroth and second moments sum rules, assuming
that the contribution of multi-hadron states is negligible below the
first excited state $M_1$, yields
\begin{equation}\label{formalbound}
\frac{1}{2 \pi i}\int_C {\mathrm d}\varepsilon\,
\theta(\varepsilon - \Delta) T(\varepsilon) 
\left( 1 - \frac{\varepsilon^2}{(E_{M_1} - E_M)^2} \right) = 
\frac{\left | \langle M |J\cdot a | B\rangle \right |^2}{4 m_B E_M} - \cdots,
\end{equation}
where the ellipsis denotes positive terms whose first derivatives at
$w=1$ are also positive.  In the next section this equation will be
used to derive a bound on $\lambda_1$.

$T(\varepsilon)$ in Eq.~(\ref{Te}) can be calculated using an OPE
\cite{BKSV}.  By taking suitable moments of $T(\varepsilon)$, it is
possible to get different sum rules that depend on the parameters of
HQET.  For the second moment sum rule, with $a=(0,0,1,0)$, this yields
\cite{BSUV,BGSUV}
\begin{equation}\label{sumrule}
\left(\frac{2w}{w + 1}\right)\left(\frac{1}{\pi}\right)
\int d\varepsilon \theta(\varepsilon - \Delta) \varepsilon^2
 {\mathrm Im}T(\varepsilon) = 
-\left(\frac{w^2-1}{3 w^2}\right) \lambda_1 + 
\overline{\Lambda}^2\left(\frac{w-1}{w}\right)^2.
\end{equation}

$T(\varepsilon)$ can be calculated to any desired order in $\alpha_s$,
thus giving perturbative corrections to the sum rules.  Here we are
concerned with the corrections to the second moment sum rule, which
are presented in the next section.

\section{results}

To calculate the $\alpha_s$ corrections, the optical theorem is used to
relate the imaginary part of $T(\varepsilon)$ to the $b \to c$ decay
rate.  At ${\mathcal O}(\alpha_s)$, the diagrams that contribute are
given in Fig.(1).
\begin{figure}[ht]
\begin{picture}(30000,10000)
\THICKLINES

\drawline\fermion[\N\REG](8000,100)[4000]
\global\Xone=\pfrontx
\global\Xtwo=\pfrontx
\global\Yone=\pfronty
\global\advance \Xone by 600
\put(\Xone,\Yone){$b$}
\drawline\gluon[\E\REG](\pbackx,\pbacky)[6]
\drawline\fermion[\N\REG](\pfrontx,\pfronty)[4000]
\global\Xthree=\pbackx
\global\Ythree=\pbacky
\global\advance\Xthree by -550
\global\advance\Ythree by -300
\put(\Xthree,\Ythree){\rule{3mm}{3mm}}
\drawline\fermion[\E\REG](\pbackx,\pbacky)[8000]
\global\advance \pbackx by 600
\put(\pbackx,\pbacky){$c$}
\global\advance\Xtwo by -250
\drawline\fermion[\N\REG](\Xtwo,\Yone)[8000]
\global\advance\pbackx by 250
\global\advance\pbacky by 250
\drawline\fermion[\E\REG](\pbackx,\pbacky)[8000]

\global\advance \Xone by 20000
\drawline\fermion[\N\REG](\Xone,\Yone)[8000]
\global\Xtwo=\pfrontx
\global\advance \Xone by 600
\put(\Xone,\Yone){$b$}
\global\Xthree=\pbackx
\global\Ythree=\pbacky
\global\advance\Xthree by -550
\global\advance\Ythree by -300
\put(\Xthree,\Ythree){\rule{3mm}{3mm}}
\drawline\fermion[\E\REG](\pbackx,\pbacky)[4000]
\drawline\gluon[\S\REG](\pbackx,\pbacky)[6]
\drawline\fermion[\E\REG](\pfrontx,\pfronty)[4000]
\global\advance \pbackx by 600
\put(\pbackx,\pbacky){$c$}
\global\advance\Xtwo by -250
\drawline\fermion[\N\REG](\Xtwo,\Yone)[8000]
\global\advance\pbackx by 250
\global\advance\pbacky by 250
\drawline\fermion[\E\REG](\pbackx,\pbacky)[8000]

\end{picture}
\caption[1]{Feynman diagrams that contribute to the $\alpha_s$ corrections of the second moment sum rule.}
\end{figure}
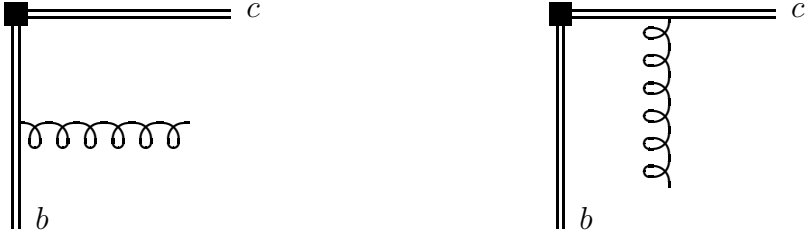
The vertex correction could in principle contribute at
${\mathcal O}(\alpha_s)$,  but is suppressed by $\alpha_s (\Lambda_{\mathrm
QCD}/m_Q)^2$, which is consistently neglected in this paper.  Therefore
we only consider the bremsstrahlung diagrams.

Using the diagrams in Fig.~(1), and expanding near zero recoil, the
leading ${\mathcal O}(\alpha_s)$ correction to the imaginary part of
$T(\varepsilon)$ is
\begin{eqnarray}\label{ImT}
\frac{1}{\pi}{\mathrm Im}T^{({\mathrm brem})}(\varepsilon_q) &=& 
\frac{8\alpha_s}{9\pi}\left[(w-1)
\frac{2 \varepsilon_q^2 + m_g^2}{\varepsilon_q^4}
\sqrt{\varepsilon_q^2 - m_g^2} \right. \\
&&\qquad-\,\left. (w-1)^2
\frac{8 \varepsilon_q^6 - 4 \varepsilon_q^4 m_g^2 - 
      \varepsilon_q^2 m_g^4 - 18 m_g^6}
{5\varepsilon_q^6\sqrt{\varepsilon_q^2 - m_g^2}}\right]
\theta(\varepsilon_q - m_g) + \cdots,\nonumber
\end{eqnarray}
where $\varepsilon_q = m_b - E_c - q_0$.  The ellipsis denotes higher
order correction in $\alpha_s$, $w-1$, and $1/m_Q$.  We have performed
the ${\mathcal O}(\alpha_s)$ calculation with a gluon mass $m_g$.  The
technique introduced in Ref.~\cite{SmithVoloshin} then allows us to obtain
the $\alpha_s^2\beta_0$ correction from this result by means of a
dispersion relation.  The partonic variable $\varepsilon_q$ is related
to the hadronic variable $\varepsilon$ by
\begin{equation}\label{epsdiff}
\varepsilon_q = \varepsilon - \overline{\Lambda}\left(\frac{w-1}{w}\right)
\end{equation}
to the order we are considering.  Rewriting Eq.~(\ref{ImT}) in terms
of $\varepsilon$, the second moment of ${\mathrm Im}T(\varepsilon)$
becomes
\eject
\begin{eqnarray}\label{fullsecondmoment}
\frac{1}{\pi}\int_C {\mathrm d}\varepsilon\,\varepsilon^2 
\theta(\varepsilon - \Delta) {\mathrm Im}T(\varepsilon) &=&
	- \frac{2}{3} (w-1) \left[ \lambda_1 - 
	\frac{4 \alpha_s(\Delta)}{3\pi}\Delta^2 - 
	\frac{2\alpha_s^2(\Delta)\beta_0}{3\pi^2}
	\Delta^2 \left(\frac{13}{6} - \ln 2\right)\right] \nonumber\\
 + \frac{4}{3}(w-1)^2&&\!\!\!\left[\left(\lambda_1 - 
	\frac{8\alpha_s(\Delta)}{15\pi}\Delta^2 -
	\frac{4\alpha_s^2(\Delta)\beta_0}{15\pi^2}
	\Delta^2  \left(\frac{187}{60} -\ln 2 \right) \right) 
	\right. \nonumber\\
&& \qquad\left. +\frac{3}{4} \left( {\overline{\Lambda}}^2 + 
	\frac{16\alpha_s(\Delta)}{9\pi}\Delta 
	\overline{\Lambda}\right)\right],
\end{eqnarray}
where $\alpha_s$ is defined in the $\overline{\mathrm MS}$ scheme.
\footnote{We disagree with the result presented in Ref.~\cite{Koyrakh}.}

The perturbative corrections to the zeroth moment sum rule were
calculated in Ref.~\cite{BLRW} to ${\mathcal O}(\alpha_s)$ and
${\mathcal O}(w-1)$.  Combining their results with
Eq.~(\ref{formalbound}) and Eq.~(\ref{fullsecondmoment}) gives
\begin{eqnarray}\label{ineq}
\left(\frac{1+w}{2w}\right)\left[ 
	1 + (w-1)\frac{8 \alpha_s(\Delta)}{9\pi}\left(\ln 4 - 
	\frac{5}{3}\right)\right]&& \nonumber\\
+ \frac{1}{\delta_1^2}\left[\frac{2}{3}
\lambda_1 (w-1) - \frac{8\alpha_s(\Delta)}{9 \pi}\Delta^2(w-1)\right] 
&& = \frac{(1+w)^2}{4w}|\xi(w)|^2 - \cdots,
\end{eqnarray}
where $\delta_1 = E_{M_1} - E_M$ is the lowest excitation energy and
again the ellipsis denotes positive terms which have positive first
derivatives at zero recoil.  We have used $|\langle D^*|J\cdot
a|B\rangle|^2 = m_B\, m_{D^*}(1+w)^2|\xi(w)|^2$.  Taking the
derivative of Eq.~(\ref{ineq}) with respect to $w$, and setting $w =
1$, gives the bound
\begin{equation}\label{bound}
\lambda_1 <
-3 \delta_1^2\left[\rho^2(\Delta) - \frac{1}{4} - 
\frac{4\alpha_s(\Delta)}{9\pi}\left(\frac{5}{3} - \ln 4\right)\right] +
\frac{4\alpha_s(\Delta)}{3\pi}\Delta^2.
\end{equation}

We define the physical slope parameter $\rho_{B \to D^*}^2$ by
\begin{equation}\label{physicalrho}
|F_{B\to D^*}(w)| = |F_{B\to D^*}(1)|[1 - \rho_{B \to D^*}^2 (w-1) + \dots].
\end{equation}
The relationship between $\rho_{B \to D^*}^2$ and $\rho^2(\mu)$ can be
computed in a model-independent way \cite{BLRW,CapriniNeubert} and is
\begin{equation}
\rho^2_{B\to D^*} = \rho^2(\mu) + 
	\frac{4 \alpha_s}{9\pi}\ln\frac{m_c^2}{\mu^2} +
	\frac{\alpha_s}{\pi}\left(\delta_{b\to D^*}^{(\alpha_s)} - 
		\frac{20}{27}\right) + 
	\frac{\overline{\Lambda}}{2m_c}\delta_{B\to D^*}^{(1/m)},
\end{equation}
where
\begin{equation}
\delta_{B\to D^*}^{(\alpha_s)} =
	\frac{2(1-z)(11+2z+11z^2)+24(2-z+z^2)z\ln z}{27(1-z)^3},
\end{equation}
and
\begin{eqnarray}
\delta_{B\to D^*}^{(1/m)} &=&
	-2\chi_1'(1) + 4\chi_3'(1) - z[2\chi_1'(1) - 4\chi_2(1)+12\chi_3'(1)]
\nonumber\\
&& - \frac{5}{6}(1+z) - \frac{4}{3}\chi_2(1) -
	\frac{1-2z+5z^2}{3(1-z)}\eta(1).
\end{eqnarray}
Here $z = m_c/m_b$ and $\delta_{B\to D^*}^{(1/m)}$ depends on the
four subleading Isgur--Wise functions $\chi_1(w)$, $\chi_2(w)$,
$\chi_3(w)$, and $\eta(w)$ \cite{Luke}; a prime denotes a
derivative with respect to $w$.  Neglecting $\delta_{B\to
D^*}^{(1/m)}$ and using $m_c = 1.4 {\mathrm\ GeV}$, $m_b=4.8
{\mathrm\ GeV}$ and $\rho_{B \to D^*}^2 = 0.91$
\cite{PDG}, we get $\rho^2(1{\mathrm\ GeV}) = 0.94$.  Choosing
$\delta_1 \approx 0.41 {\mathrm\ GeV} = M_{D_1} - M_{D^*}$ gives
\begin{equation}\label{pertbound}
\lambda_1 < -0.15{\mathrm\ GeV}^2.
\end{equation}
This should be contrasted with the bound without perturbative
corrections \cite{BGSUV}, $\lambda_1~<~-0.33{\mathrm\ GeV}^2$,
obtained with the same choice of $\delta_1$.

\section{discussion}

The bound on $\lambda_1$ derived at tree level is significantly
weakened by perturbative corrections and depends sensitively on
$\delta_1$ and $\Delta$.  In Table \ref{boundtable}, we present
various bounds on $\lambda_1$ obtained for different values of the
parameters in Eq.~(\ref{bound}).

\begin{table}[bt]
\begin{tabular}{c||cc|cc} 
&  \multicolumn{2}{c|}{$\Delta = 1\,$GeV}  
  &  \multicolumn{2}{c}{$\Delta = 1.5\,$GeV}  \\
&  $\delta_1 = 0.41\,$GeV  &  $\delta_1 = 0.50\,$GeV   
&  $\delta_1 = 0.41\,$GeV  &  $\delta_1 = 0.50\,$GeV   \\  \hline   
~~Tree~~~~  &  $-0.33$  &  $-0.50$  &  $-0.33$  &  $-0.50$  \\
~~Order $\alpha_s$~~~~  &  $-0.15$  &  $-0.32$  &  $-0.03$  &  $-0.21$
\end{tabular} \vskip6pt
\caption[]{Upper bound on the HQET parameter $\lambda_1$ at tree level 
and order $\alpha_s$.  The bound is extremely sensitive to changes in
the values of the cutoff $\Delta$ and the minimal excitation energy
$\delta_1 = E_{M_1} - E_M$.
\label{boundtable}}
\end{table}

The value of $\rho^2$ is not well known.  To estimate the $1/m_Q$
effects, the subleading Isgur--Wise functions must be known near zero
recoil.  QCD sum rule predictions \cite{NLN} give approximately
$\chi_1'(1) = 0.3,\ \chi_2(1) = -0.04,\ \chi_3'(1) = 0.02$ and
$\eta(1) = 0.6$.  Using these values, which have large uncertainties,
and $\overline{\Lambda} = 0.4\ {\mathrm GeV}$ we find that
$\rho^2(\Delta)$ increases by about $0.3$.  Adding this contribution
to $\rho^2$, with $\Delta =1\ {\mathrm GeV}$ and $\delta_1=0.41\
{\mathrm GeV}$, the bound becomes $\lambda_1 < -0.30{\mathrm\
GeV}^2$, which provides an indication of the size of these effects.

It is always possible to absorb some perturbative corrections into the
definition of $\lambda_1$ and $\overline{\Lambda}$.  This may result
in a perturbative series that is better behaved.  For instance, we may
redefine $\lambda_1$ by
\begin{equation}\label{lambda1shift}
\lambda_1 \to \lambda'_1(\Delta) = \lambda_1 - 
   \frac{4 \alpha_s(\Delta)}{3 \pi} \Delta^2 -
   \frac{2 \alpha_s^2(\Delta)\beta_0}{3 \pi^2} \Delta^2 
	\left(\frac{13}{6} - \ln 2\right),
\end{equation}
which will remove the perturbative corrections to the second moment
sum rule at order $w-1$.  The redefinition also removes the
corrections to the zero recoil sum rule in Eq.~(19) of
Ref.~\cite{KLWG}, and is consistent with the suggested definition in
Ref.~\cite{BSUV}.  The bound in this case, with $\delta_1 =
0.41{\mathrm\ GeV}$, becomes
\begin{equation}\label{redefbound}
\lambda'_1(\Delta = 1{\mathrm\ GeV}) < - 0.34 {\mathrm\ GeV}^2.
\end{equation}
However, we observe from Eq.~(\ref{fullsecondmoment}) that the
redefinition of $\lambda'_1$ does not completely remove the
perturbative correction at order $(w-1)^2$.

Similarly, one can redefine $\overline{\Lambda}$ by absorbing the
perturbative corrections.  To do this, one must use the terms
proportional to $\overline{\Lambda}$ and $\overline{\Lambda}^2$ in
Eq.~(\ref{fullsecondmoment}) to form a complete square.  This
results in the following redefinition of $\overline{\Lambda}$
\begin{equation}\label{Lambdabarshift}
\overline{\Lambda} \to \overline{\Lambda}'(\Delta) = \overline{\Lambda} + 
		\frac{8 \alpha_s(\Delta)}{9 \pi} \Delta.
\end{equation}
This redefinition does not, however, remove the perturbative
corrections to the Voloshin sum rule of Ref.~\cite{BLRW}.  It
is also different from the $\overline{\Lambda}$ defined by $m_B =
\overline{m}_b(\mu) + \overline{\Lambda}(\mu) + \cdots$ using the
$\overline{\mathrm MS}$ quark mass, which also appears in the
literature.  Thus, it is not possible to completely remove the
perturbative corrections by redefining $\lambda_1$ and
$\overline{\Lambda}$.

\acknowledgments We wish to thank M. Wise and Z. Ligeti for many useful
conversations.  We would also like to thank M. Gremm, A. Kapustin, and
I. Stewart for helpful comments.

{\tighten

}

\end{document}